\documentclass[aps,twocolumn]{revtex4}

\usepackage{graphicx}
\newcommand{\SLASH}[1]{/\!\!\! #1}

\def\be{\begin{eqnarray} &&}

\def\ee{\end{eqnarray}}

\usepackage{graphicx} %%Include figure files
\usepackage{dcolumn} %%Align table columns on decimal point
\usepackage{bm} %%bold math
\usepackage{epsfig}

\begin{document}

\title{Modeling electromagnetic form factors of light and heavy pseudoscalar mesons}
\author{B.~El-Bennich$^{1,2}$, J. P. B. C.~de Melo$^3$, B.~Loiseau$^1$, J.-P.~Dedonder$^1$,
 and
T.~Frederico$^4$}
\address{$^1$ Laboratoire de Physique Nucl\'eaire et de Hautes \'Energies,  Groupe Th\'eorie, \\
 CNRS, IN2P3 \& Universit\'es Pierre et Marie Curie et Paris-Diderot, 75005 Paris, France}
 \address{$^2$ Physics Division, Argonne National Laboratory, Argonne, IL 60439, USA}
\address{$^3$Laborat\'orio de F\'\i sica Te\'orica e Computa\c{c}\~ao
Cient\'\i fica, Centro de Ci\^encias Exatas e Tecnol\'ogicas,
Universidade Cruzeiro do Sul, 08060-700  and
Instituto de F\'isica Te\'orica, 01405-900, S\~ao Paulo, Brazil}
\address{$^4$ Dep. de F\'isica, Instituto
Tecnol\'ogico de Aeron\'autica, 12.228-900 S\~ao Jos\'e dos Campos, Brazil}

\begin{abstract}
The electromagnetic form factors of light and heavy pseudoscalar
mesons are calculated within two covariant constituent-quark models,
 {\em viz.\/},  a light-front and a dispersion relation approach. We investigate the details and physical origins of the model dependence of
 various hadronic observables: the weak decay constant, the charge radius and the elastic electromagnetic form factor.
\end{abstract}


\maketitle

\section{Introduction}
\vspace*{-1mm}

The light-front quantum field theory is discussed by N.~Bogoliubov~{\em et al.\/} in Ref.~\cite{Bogoliubov} and an early application to study
hadronic quark bound states may be found in Refs.~\cite{Terentev1,Terentev2}. A self-consistent relativistic treatment of the quark spins
can be performed within the light-front quark model, which allows for a calculation of the reference-frame independent partonic contribution
to the form factor. The non-partonic contribution cannot be generally calculated, however it can be eliminated for space-like momentum
transfers by an appropriate choice of reference frame. Therefore, the partonic contribution obtained in this specific reference frame
yields the full form factor.

The light-meson sector, which includes the pion and kaon, allows us to test QCD hypotheses on the subatomic structure of hadrons at low
and intermediate energies. Yet, many aspects of quantum field theory on the light-front and its application to bound-state systems
give rise to various open questions; for example, the problems of regularization and renormalization on the light-front.
For the pion and kaon, the light-front constituent quark model (LFCQM) has been studied~\cite{Pacheco99,Pacheco2002} to describe
recent experimental data~\cite{Baldini,TJAB,Horn,Tadevosyan}.

 The description of these bound states using dispersion relations
(DR) was first developed in ~\cite{Anisovich92}.  It was applied to
calculate  light mesons form factors~\cite{Anisovich95,Anisovich97}
and used to elucidate long-distance effects
~\cite{melikhov1,melikhov2} in weak decays of heavy mesons. More
recently the interplay between perturbative and nonperturbative
regions in the pion electromagnetic form factor was discussed within
the DR approach \cite{braguta08}, where it was pointed out the
dominance  of the nonperturbative contribution up to rather high
values of momentum transfers, going beyond a previous analysis
performed up to intermediate values of momentum transfers within
light-cone sum rule approach~\cite{braum00}.

The DR approach is based on a consistent treatment of the
two-particle singularities which arise in the triangle diagrams
describing elastic as well as inelastic meson-transition amplitudes.
In this dispersive approach, these amplitudes are given by
relativistic spectral integrals over the mass variables in terms of
Bethe-Salpeter amplitudes of the mesons and spectral densities of
the corresponding triangle diagrams. Thus, elastic form factors are
described by double spectral representations.

In this paper, we analyze the model dependence which, due to vertex
functions as well as the choice of a constituent quark mass, usually
arises in quark model calculations of elastic form factors. In the
light sector, we compute the pion and kaon electric form factors for
which experimental data is available; we then use our models to make
predictions for the charmed sector.   For simplicity we will assume
point-like  constituent quarks, as our model assumptions are made
directly on the explicit form of wave functions and vertices.
However, the constituent quark may have a structure that depends on
the dynamical model of the bound state (see e.g. \cite{lucha06}).

\section{Electromagnetic form factors }
\vspace*{-1mm}

The space-like electromagnetic form factor of a pseudoscalar meson
with mass $M$ is generally given by the covariant expression
\begin{equation}
  \langle P(p')  |  J^\mathrm{em}_\mu  | P(p)\rangle = (p + p')_\mu F^\mathrm{em}(q^2) \ ,
\end{equation}
with $p^2=p'^2=M^2$ and the four-momentum transfer $q=p'-p$,
$q^2<0$. The electromagnetic current  is $J^\mathrm{em}_\mu =\bar
q(0)\gamma_\mu q(0)$  where $q(0)$ denotes a current quark. Here,
$F^{\mathrm{em}}(q^2)$ describes the virtual photon emission
(absorption) amplitude by the composite state of a quark of charge
$e_1$ and an antiquark of charge $e_2$. This form factor depends on
the three independent Lorentz invariants $p^2$, $p'^2$ and $q^2$.
The constituent quark amplitude of the electromagnetic form factor
is assumed to have the following structure
\begin{equation}
  \langle Q(p')  | \bar q(0) \gamma_\mu q(0) | Q(p)
\rangle = \bar Q(p')\gamma_\mu Q(p)\, \xi_c(q^2) \ ,
  \label{const}
\end{equation}
where $Q(p)$ represents the constituent quark. The form factor is
normalized such that $F^\mathrm{em}(0)= e_1+e_2$ and we neglect the
anomalous magnetic moment of the constituent quark in
Eq.~(\ref{const}). The function $\xi_c(q^2)$ describes a constituent
quark transition form factor. Since the quark model is not formally
derived from QCD, it is unknown. In the following we assume that
$\xi_c\simeq e_c=e_1$ or $e_2$ owing to the fact that constituent
quarks behave like bare Dirac particles~\cite{Weinberg:1990xm}.

\section{Light-Front Constituent Quark Model}

The electromagnetic form factor  for pseudoscalar particles is calculated in the impulse approximation, using the Mandelstam formula
\begin{eqnarray}
 F^\mathrm{em}(q^2) & = & \frac{e_1 N_c}{(p+p^{\prime})^{\mu}} \int \frac{d^4k}{(2\pi)^4} \mathrm{Tr} \left[ \gamma^5 S_1(k-p^{\prime}) \ \times \right.
\nonumber \\ && \hspace*{-1.6cm} \left.  \gamma^{\mu}S_1(k-p)\gamma^5 S_2(k) \right] \Lambda(k,p) \Lambda(k,p^{\prime}) + [1 \leftrightarrow 2]\ ,
\label{formfactor}
\end{eqnarray}
where $S_i(p)$ is the Feynman propagator of quark $i$ with constituent mass $m_i$, $N_c$ the number of colors, and $\Lambda(k,p)$ is our hadron-quark vertex function model. The bracket $[1\leftrightarrow 2]$ is a shortcut for
the subprocess of constituent quark $2$ interacting with the photon.

The momentum component Bethe-Salpeter (BS) vertex model is
chosen such that it regularizes the amplitude of the photo-absorption
process and constructs a light-front valence wave
function~\cite{Pacheco99,Pacheco2002,Ji2001}. In the present study, we use two different
models for the vertex functions taken from previous work, one of which has a non-symmetrical form under the exchange of the
momentum of the quark and antiquark~\cite{Pacheco99},
\begin{equation}
 \Lambda(k,p)=\frac{N}{( p-k)^2-m^2_R+i \varepsilon }\, , \label{vnsy}
\end{equation}
while the other has a symmetric form \cite{Pacheco2002},
\begin{equation}
 \Lambda(k,p)= \left[ \frac{N}{ k^2-m^2_R+i \varepsilon } + \frac{N}{ ( p-k)^2-m^2_R+i \varepsilon }\right] \, .
\label{vsy}
\end{equation}
The normalization constant $N$ is obtained from $F^\mathrm{em}(0)=e_1+e_2$.

The light-front constituent quark model for hadrons used here is based on quantum field theory and our ansatz comes with the choice of the BS
vertex and point-like quarks.  It is worthwhile to mention that the present LFCQM reproduces in its full complexity a covariant calculation.
Technically, the use of the Drell-Yan frame (longitudinal momentum transfer $q^+=q^0+q^3=0$) and the projection on to the light-front
(performing the integration on the minus momentum component analytically in the Mandelstam formula) simplifies drastically the
computation of the form factor. This approach has been applied to the pion~\cite{Pacheco99,Pacheco2002,Ji2001}, the kaon~\cite{Fabiano}
as well as to the $\rho$~\cite{Pacheco97}.

The use of light-front variables in the evaluation of the Mandelstam formula, with the corresponding projection on the light-front hypersurface
through the $k^-=k^0-k^3$ integration, has its subtleties. Covariance of the starting expression in Eq.~(\ref{formfactor}), which corresponds to a
frame independent form factor, is an important property to be checked in the final results. In some cases, as for vector mesons in the present model,
it is necessary to perform a careful analysis of the light-front calculation of the photo-absorption amplitude which accounts for pair terms or
Z-diagrams that survive in the Drell-Yan frame (see, e.g., Ref.~\cite{Pacheco98}). The Z-diagram vanishes in the calculation of the plus
component of the electromagnetic current $(j^{+})$ with $q^+=0$ for pseudoscalar mesons with $\gamma5$ coupling
and the momentum component of the BS vertices from Eqs.~(\ref{vnsy}) and (\ref{vsy}).

Therefore, we adopt the plus-component convention of the electromagnetic current in the Breit-frame with the Drell-Yan condition ($q^+=0$) to
compute the form factor of pseudoscalar mesons. The $k^-$ integration is performed analytically by evaluating the residues in Eq.~(\ref{formfactor}).
In the case of the non-symmetric vertex (\ref{vnsy}), the resulting expression for the electromagnetic form factor is:
 \begin{eqnarray}
 F^\mathrm{em}(q^2) & = &  e_1 \frac{N^2 N_c}{(p^{+} + p^{\prime+ })} \int \frac{d^{2} k_{\perp} d k^{+} }{4\pi^3}
 \frac{  \mathrm{Tr}^+_1[ \ \ ]|_{k^-=k^-_{2\mathrm{on}}} }{k^+(p^+-k^+)^2 }   \nonumber \\
 & & \hspace*{-1.6cm} \times\ \frac{\Lambda(k,p)|_{k^-=k^-_{2\mathrm{on}}} \Lambda(k,p^{\prime})|_{k^-=k^-_{2\mathrm{on}}}}
  {\left(p^- - k^-_{2\mathrm{on}} - (p-k)^-_{1\mathrm{on}}\right) \left(p^{\prime -} - k^-_{2\mathrm{on}} - (p^\prime-k)^-_{1\mathrm{on}} \right) } \nonumber \\
 &  + &  [1 \leftrightarrow 2] \  , \label{ffactor+}
\end{eqnarray}
where the on-minus-shell momentum is $k^-_{1\mathrm{on}}=(k^2_\perp+m_1^2)/k^+$ ($k^-_{2\mathrm{on}}=(k^2_\perp+m_2^2)/k^+$) and
the Dirac trace from Eq.~(\ref{formfactor}) for $\mu=+$  (good component of the current) is
\begin{eqnarray}
\mathrm{Tr}_1^+[\ \ ] &=& -4\left[ (p^+-k^+)\left(2m_1m_2+(p+p')\cdot k -2k^2\right)
\right. \nonumber \\ && \left.
-k^+\left((p'-k)\cdot(p-k)- m_1^2\right)\right].
\end{eqnarray}
The electromagnetic form factor written for the non-symmetric vertex (\ref{vnsy}) in terms of the Bjorken momentum fraction $x=k^+/p^+$ is given by
 \begin{eqnarray}
 F^\mathrm{em}(q^2) & = &  e_1 \frac{N^2 N_c} {8\pi^3} \int \frac{d^{2} k_{\perp} d x}{x(1-x)^4}    \nonumber \\
&  & \hspace{-1.5cm} \times \  \frac{\xi(x,\vec k_\perp,\vec q_\perp, m_1,m_2)}{(M^2-M^{\prime 2}_{0}) (M^2-M^{\prime 2}(m_2,m_R))}
 \nonumber \\
& &  \hspace{-1.5cm} \times \ \frac{1}{(M^2-M^2_{0}) (M^2-M^2(m_2,m_R))} + [1 \leftrightarrow 2] \, ,
\label{ffactor+a}
\end{eqnarray}
where $\xi(x,\vec k_\perp,\vec q_\perp, m_1,m_2)= (p^+)^{-1} \mathrm{Tr}^+_1[\ \ ]|_{k^-=k^-_{2\mathrm{on}}}$, $0<x<1$ and
the free two-quark mass is $M^2_{0}=M^2(m_2,m_1)$, with
\begin{equation}
M^2(m_i,m_R)=\frac{k^2_\perp+m^2_i}{x}+ \frac{(p-k)^2_\perp+m^2_R}{1-x}-p^2_\perp \ .
\label{fupi2}
\end{equation}
Analogous expressions follow for $M^{\prime 2}_{0}$ and $M^{\prime 2}(m_2,m_R)$ with $ p^{2}_\perp$ replaced by $ p^{\prime
2}_\perp$. The calculation of the electromagnetic form factor with the symmetrical vertex (\ref{vsy}) can be done following the steps presented
above. For the interested reader, the derivation of the form factor for the pion in this model is performed in detail in Ref.~\cite{Pacheco2002}.

The  mass parameters in these models are limited by $m_1+m_2>M$, $m_1+m_R>M$ and $m_2+m_R>M$ due to the unphysical continuum
threshold as seen in the denominator of Eq.~(\ref{ffactor+a}). Quark confinement does not allow for a scattering cut, but since in the present model
the meson is as a real bound state, {\em i.e.\/} the conditions $m_1+m_2>M$ and $m_i+m_R> M$ are satisfied, the cut is harmless.
However, if the meson is a weakly bound composite particle, the form factor will be very sensitive to changes in the constituent quark mass.
This is certainly not the case for pions and kaons, as they are Goldstone bosons and strongly bound, yet this sensitivity appears in the case of heavy pseudoscalars with small binding energies in nonconfining models (see, {\em e.g.\/}, Ref.~\cite{pauli2001,pacheco2006}). In particular, our numerical
results for $D^+$ will exemplify the strong dependence of the heavy pseudoscalar electromagnetic form factor on the constituent masses, whereas this effect is minor for confining models like the DR approach.

\section{Dispersion relation approach}

\begin{figure}[t!]
\begin{center}
 \epsfig{figure=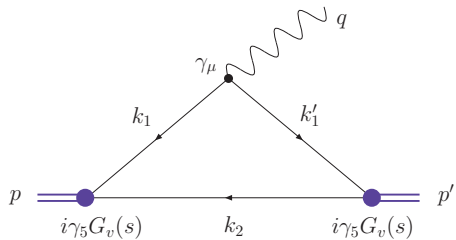,scale=0.6,angle=0} \vspace*{-5mm}
\caption{The electromagnetic form factor in the impulse approximation is obtained from the triangle diagram. The vertices and momenta
depicted in the diagram represent the momentum assignment in the DR approach.}
\label{triangle}
\end{center}\vspace*{-3mm}
\end{figure}

The  amplitude is obtained from the triangle diagram shown in Fig.~\ref{triangle} where the kinematical variables are displayed. The on-shell meson momenta are $p^2=M^2$ and $p'^2=M^2$. The triangle diagram may be calculated in various ways --- we choose to put the constituent quarks on-mass shell while keeping the external momenta off-shell with
\begin{equation}
\label{qoffshell}
\tilde p^2=s\ ,\quad  \tilde p'^2=s'\ ,\quad (\tilde p'-\tilde p)^2=q^2\ .
\end{equation}
Note, however, that  $\tilde p'- \tilde p = \tilde q \neq q$.

In the DR approach, the electromagnetic form factors
$F^\mathrm{em}(q^2)$ are expressed by a double spectral
representation %~\cite{melikhov1,melikhov2},
(see references
~\cite{Anisovich92,Anisovich95,Anisovich97,melikhov1,melikhov2}),
\begin{eqnarray}
\label{fplusmoins} F^\mathrm{em} (q^2) & = & e_1
\int \frac{ds\ G_v(s)}{\pi(s-M^2)}\ \frac{ds'\ G_v(s')}{\pi(s'-M^2)}\nonumber \\
 &  \times & \ \Delta (s,s',q^2|m_1,m_2)\ +\ [1\leftrightarrow 2 ] \ .
\end{eqnarray}
We apply the Landau-Cutkosky rules to calculate the double spectral density $\Delta_V$:  we place all internal particles on their mass shell,
$k^2_1 = m^2_1$, $k'^2_1 = m^2_1$, $k_2^2 =m^2_2$ but take the variables $p^2$ and $p'^2$ off-shell. The double spectral density is then
derived from calculation of the triangle diagram
\begin{widetext}
\begin{eqnarray}
\label{doubles}
  2\, \tilde p_\mu (q) \Delta (s,s',q^2 | m_1,m_2) & =
&\displaystyle\frac{1}{8\pi}\int d^4k_1d^4k_1'd^4k_2\ \delta(k_1^2-m_1^2)
     \delta(k_1'^2-m_1^2) \delta(k_2^2-m_2^2)  \nonumber \\
  & & \hspace*{-1.5cm} \times \
\delta(\tilde p-k_1-k_2)\delta(\tilde p'-k_1'-k_2)\  \mathrm{Tr}\left[-(\SLASH k_1'+m_1)\gamma^\mu(\SLASH k_1+m_1)i
       \gamma^5 (m_2-\SLASH k_2)i\gamma^5 \right]  ,   \hspace{4mm}
\end{eqnarray}
\end{widetext}
with $\tilde p_\mu (q) = \tilde p_\mu - \frac{\tilde p\cdot q}{q^2}\ q_\mu$
which ensures the Ward identity and thus charge conservation $F^\mathrm{em}(0)=e_1+e_2$ with the proper vertex normalization.

The vertex of the pseudoscalar meson in the constituent quark picture has the structure
\begin{equation}
\label{vertexs}
\frac{\bar Q^{a}(k_1,m_1)i\gamma^5 Q^{a}(-k_2,m_2)}{\sqrt{N_c}}\ G_v(s)\ ,
\end{equation}
where $Q^{a}(k_1,m_1)$ represents the spinor state of the constituent quark of color $a$ and $N_c$  the number of colors. The bound state vertex
function may be related to the mesonic BS amplitude by
\begin{equation}
  \phi (s) = \frac{G_v(s)}{s-M^2} \ .
\end{equation}
For a confining potential, the pole at $s = M^2$ should appear in the physical region for $\phi(s)$ at $s = M^2 > (m_1+m_2)^2$.
This is, however, not the case; as well known from the behavior of the bound-state wave function in an harmonic oscillator potential,
$\phi (s)$ is a smooth exponential function of $s \geq (m_1 + m_2)^2$. This means that the would-be pole in $\phi (s)$ at
$s = M^2$ is completely blurred out by the interaction and it is therefore more appropriate to analyze the meson form factors in
terms of $\phi (s)$ rather than $G_v(s)$.

For a pseudoscalar meson in the dispersion approach of the constituent quark model, the BS amplitude $\phi(s)$ which accounts for
soft constituent quark rescattering is given by \cite{melikhov1}
\begin{equation}
\label{phi} \phi(s)= \frac{\pi}{\sqrt{2}}\ \sqrt{\frac{s^2-(m_1^2-m_2^2)^2}{s-(m_1-m_2)^2}} \ \frac{w(k)}{s^{3/4}} \ .
\end{equation}
The dynamical factor multiplying $w(k)$ stems from the loop diagram associated with the meson-vertex normalization.
The modulus of the center of mass momentum is
\begin{eqnarray}
\label{smallk} k = \sqrt{\frac{(s +m_1^2-m_2^2)^2 -4 s m_1^2}{4 s}} \ .
\end{eqnarray}

It can be shown \cite{melikhov1} that the vertex normalization of $G_v(s)=\phi(s) (s-M^2)$, which describes soft constituent rescattering,
reduces to a simple normalization of the wave function
\begin{equation}
\label{normw}
  \int_0^\infty w^2(k)k^2\ dk=1\ .
\end{equation}
A heuristic choice must be made for $w(k)$. For phenomena that are predominantly governed by infrared mass scales, it is sensible in
the case of heavy mesons to choose a BS amplitude parameterized by a function whose support is in the infrared; for instance functions
of Gaussian form,
\begin{equation}
\label{wk}
w(k)= N \exp\left(-4\nu k^2/\mu^2\right)\ ,
\end{equation}
where the reduced mass of the quark-antiquark pair is $\mu=m_1m_2/(m_1+m_2)$ and $N$ a normalization constant. For light
pseudoscalar mesons, dynamical chiral symmetry and the Ward-Takahashi identity are of great importance for their structure
and properties. In non-perturbative approaches, using for example Schwinger-Dyson equations which incorporate these QCD features,
appropriate parameterizations for the pion or kaon BS amplitudes are derived \cite{cdroberts}. In this work, we employ power-law wave
functions of the form
\begin{equation}
w_n(k)= \frac{N}{(1+ \nu (k/\mu)^2 )^n}\ , \quad n \geq  2 \ ,
\end{equation}
inspired by the form suggested in Ref.~\cite{schlumpf94} from the analysis of the ultraviolet physics of QCD.

The size parameter $\nu$ is to be determined from experimental and theoretical considerations. On the experimental side, on can either
choose to constrain the vertex functions, or equivalently the BS amplitude, by the weak decay constant or by the electric charge radius.
It is worthwhile to observe that both wave function models, Gaussian and power-law, do not present the singularity problem brought
by the scattering cut due to changes the constituent quark masses, differently from the present LFCQM.

\section{Numerical Results}

\begin{figure}[t!]

\begin{center}
\vspace*{-2mm}
\epsfig{figure=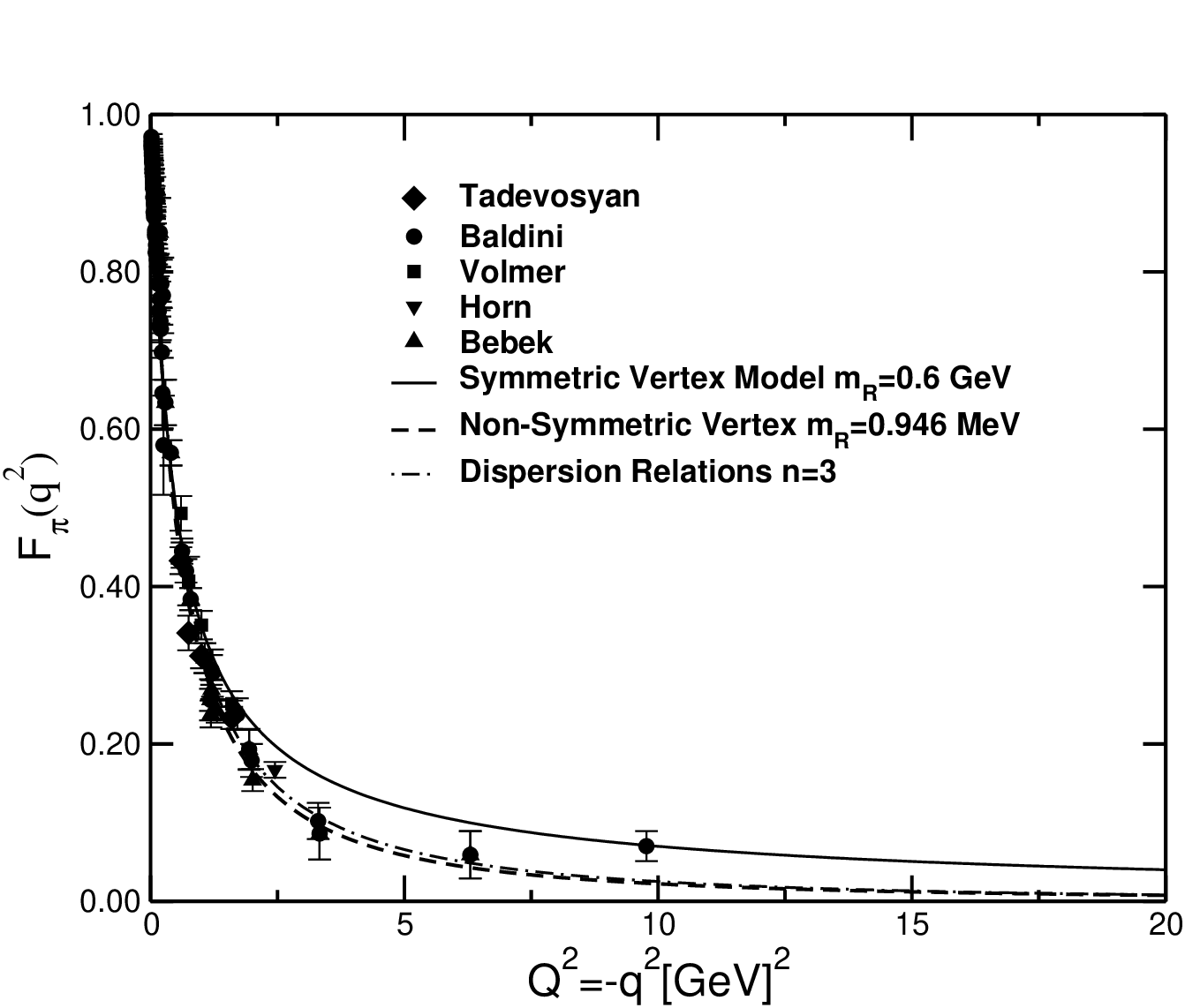,scale=0.37,angle=0}
\caption{The pion electormagnetic form factor in the LFCQM  and DR approaches compared with
experimental data from Ref.~\cite{Baldini,TJAB,Horn,Tadevosyan}.
The model and respective parameters are explained in the legend.}
%%from
%%new experimental data~\cite{Volmer2001}~(square),
%%\cite{Frascati2001}~(circle),
%%\cite{Horn2006}~(triangle up) and
%%\cite{Tadevosyan2006}~(triangle down ).
\label{fig2}
\end{center} \vspace*{-6mm}
\end{figure}

The electromagnetic form factor of light and heavy pseudoscalar mesons are calculated with the two covariant constituent quark models.
In the LFCQM, the quarks are in a bound state while in the DR model the wave functions corresponds to confined quarks with Gaussian and power-law
forms. The shape of these functions is obtained from a  fit to the experimental values of $f_\pi, f_K$ and $f_D$ or the charge radius (when available)
of the regulator mass $m_R$ in the LFCQM and size parameter $\nu$ in the DR approach. The light constituent quark masses $m_u=m_d$
are allowed to take the values 0.22 and 0.25~GeV (and 0.28~GeV in case of the $D^+$). The strange and charm constituent masses are fixed to
$m_s=$~0.508 GeV and $m_c=$~1.623~GeV, respectively (see Ref.~\cite{pachecoprd02}).

The results for the pion electromagnetic form factor compared to experimental data \cite{Baldini,TJAB,Horn} are shown in Fig.~\ref{fig2}.
The quark masses used in these calculations are $m_u=m_d=0.22$~GeV in both the LFCQM and DR approaches. The pion charge radius is used
to fit the parameters $m_R$ and $\nu$ (power-law wave function with $n=3$). The regulator mass is $m_R=0.946$~GeV~\cite{Pacheco99} for the
non-symmetric vertex and $0.546$~GeV~\cite{Pacheco2002} the symmetric vertex. In this case, we show results only for
the power-law wave function with $n=3$ ($\nu=0.088$) in the DR approach because the Gaussian vertex produces a form factor strongly damped with
increasing $q^2$, which violates the asymptotic QCD result predicting a $q^{-2}$ falloff.
The pion electromagnetic form factor for the power-law wave function and non-symmetric vertex are very similar,
while the symmetrical model has a longer tail. These results indicate that the electromagnetic form factor of the pion as a strongly bound system
does not particularly distinguish between models with a scattering cut. The high momentum experimental data appear to favor the more complex
structure of the symmetrical vertex. It is expected that in this region the details of the regulator structure are crucial, and moreover a symmetric form
of the vertex is natural.

We remind that the tail of the momentum component of the pion light-front wave function corresponding to the non-symmetric  \cite{Pacheco99} and
to the symmetric vertex \cite{Pacheco2002} both decrease as $\sim k_\perp^{-4}$, which can also be verified by inspecting Eq.~(\ref{ffactor+a}).
The power-law wave function model with $n=3$ also displays a tail that falls off with $\sim k_\perp^{-4}$. Thus, in the space-like region up to
10~GeV$^2$,  we find that the pion electromagnetic form factor favors the power-law damping of the wave function with a $\sim k_\perp^{-4}$ tail.

%%Fig.3
\begin{figure}[t!]
\begin{center}\epsfig{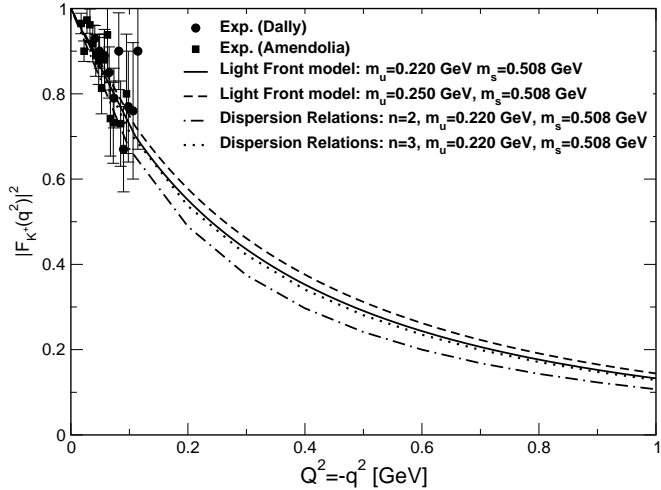}
\caption{The kaon electromagnetic form factor in the  LFCQM  and DR approaches compared with experimental data from
Refs.~\cite{Dally,Amendolia}. The models and respective parameters are explained in the legend}
\label{fig3}
\end{center}\vspace*{-6mm}
\end{figure}

The electromagnetic form factor of the $K^+$ obtained in both models are compared with experimental data \cite{Dally,Amendolia} in Fig.~\ref{fig3}.
The quark masses used in these calculations are $m_u=m_d=0.22$~GeV and $m_s=0.508$~GeV in both the non-symmetric vertex and DR approaches. In order to study the mass dependence of the form factor, we also give results for a calculation with $m_u=m_d=0.25$~GeV in LFCQM case. The regulator mass is fixed to the pion's value $m_R=0.946$~GeV  for the non-symmetric vertex which corresponds to a reasonable kaon charge radius~\cite{Fabiano}. We observe in Fig.~\ref{fig3} that the change of 0.03~GeV in the constituent quark mass in is not important as it is much smaller than the kaon binding energy of about 0.2~GeV. A similar finding can be reported for the DR approach, where the size parameter $\mu$ of the power-law wave function with
$n=2$ is fitted ($\nu=0.061$) to reproduce the kaon charge radius and is readjusted for $n=3$ ($\nu=0.158$).  Again, we observe that the non-symmetric vertex as well as the power-law model essentially yield the same form factor once the radius is reproduced.

%%Fig.4
\begin{figure}[t!]
\begin{center}
\epsfig{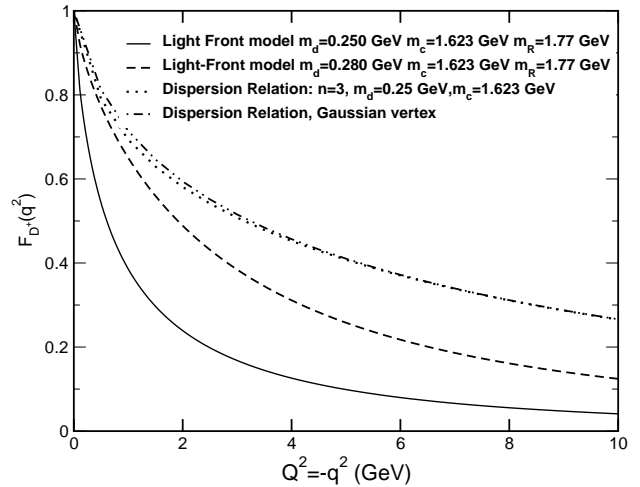}
\caption{Predictions for the $D^+$ electromagnetic form factor in the LFCQM and DR approach.}
\label{fig4}
\end{center}\vspace*{-4mm}
\end{figure}

Our numerical results for $D^+$ are shown in Fig.~\ref{fig4} for  the DR approach and LFCQM. The size parameter of the Gaussian and power-law
($n=3$) models are fitted to $f_D$ for $m_d=0.25$~GeV. Up to 10~GeV$^2$ there is no sizable difference between the two calculations.
This is reasonable as $m_c$ sets one large scale whereas the momentum transfer goes only up to 3~GeV, which is not large enough to
discriminate between the models. The regulator mass of 1.77~GeV in the non-symmetric vertex model is chosen so as to  give a sample
of the LFCQM results. Here, we want to illustrate the constituent quark mass sensitivity when the symmetric
and non-symmetric vertices are used. We remind that the mass parameters in these models are limited by $m_1+m_2>M$, $m_1+m_R>M$ and
$m_2+m_R>M$ due to the unphysical continuum threshold as evident from the denominator in Eq. (\ref{ffactor+a}). The model binding energy of
$D^+$ is only a few tenths of MeV, as can be inferred  from the $D^+$ and quark masses. Since the $D^+$ meson mass is fixed to the experimental
value of 1.8694~GeV and $m_c=1.623$~GeV, the light quark mass $m_d$ must be larger than the value 0.22~GeV we used for the pion and kaon. The increase of the light quark constituent mass from 0.25 to 0.28~GeV corresponds to a large increase in binding and consequently to a decrease of the radius, as seen in Fig.~\ref{fig4}.

\section{Conclusion}

We have concentrated on the model dependence of elastic electromagnetic form factors of light and heavy pseudoscalar mesons.
Two different cases were studied by changing the form of the vertex or wave function and by varying the constituent quark masses. In
the light sector, we computed the pion and kaon electric form factors for which experimental data is available; we then adjusted our models to
make predictions for the charmed sector. Two ansatzes have been studied in the LFCQM which involve a symmetric and a non-symmetric
vertex form. In the DR approach, Gaussian and power-law wave functions were used in the calculations.

The form factors of the pion and kaon, both being strongly bound systems, are sensitive to the short-range part of the vertex functions.
Therefore, the tail of the momentum component of the light-front wave functions for the non-symmetric~\cite{Pacheco99}
and symmetric vertex~\cite{Pacheco2002} as well as the power-law model ($n=3$), all decreasing as $\sim k_\perp^{-4}$,
is essential to reproduce the pion space-like data up to 10~GeV$^2$. Moreover, as a consequence of the strong binding of the
quarks, changes in the constituent mass of about ten percent are not important for the elastic form factor of light mesons, provided the
charge radius (or decay constant) is accordingly refitted. We have exemplified this in the kaon calculation. The light quark constituent
masses are relevant for low momentum transfers of the order $m^2_{u,d,s}$ around $0.05-0.3$GeV$^2$, which is dominated by the
charge radius and consequently compensated by fitting the size or regulator mass parameters. Therefore, in qualitative agreement
with QCD scaling laws in the ultraviolet limit, a power-law form of the wave function and our choice of vertices are reasonable
in reproducing the form factor data once the charge radius is fitted.

 A recent calculation of the pion form factor within the dispersion
relation approach including  $\mathcal {O}(\alpha_S)$ QCD
perturbative contribution \cite{braguta08}, showed in a model
independent way that the nonperturbative part stays above 50\% for
$Q^2\leq 20$~GeV$^2$. The nonperturbative part is compatible with a
pion wave function close to the asymptotic one, with a momentum
scale given by the effective continuum threshold that determines to
a great extent their results for the form factor. The value of the
threshold is extracted from the experimental pion decay constant.
Similarly in our analysis the size or regulator mass parameters are
fixed by the experimental values of the weak decay constants within
models that have momentum tails decreasing with $\sim k_\perp^{-4}$,
which does not leave not too much freedom for the pion form factor
results below 20~GeV$^2$, as we have shown. In that respect our
calculation and \cite{braguta08} show the dominance of the scale
given by the pion decay constant that dials the vertex or wave
function (decaying as a power law)  in the pion form factor up to a
fairly large value of momentum transfers.

The electromagnetic form factor of pseudoscalar mesons in the heavy-light sector is strongly sensitive to the infrared  behavior of QCD,
compared to the case of light mesons. The relevance of soft physics in this sector is supported by our calculations of the elastic $D$ meson
form factor both within the dispersion relation and light-front approaches. Once the size parameter of the Gaussian and power-law wave
functions are adjusted to reproduce the decay constant, the elastic form factor is insensitive to the vertex model in the space-like
region up to 10~GeV$^2$. Therefore, the large-momentum tail of the wave function is not important for these results.  We presume that
the asymptotic behavior will be important only for $q^2\gg m^2_c$ beyond the results we presented. The importance of soft physics may
also be appreciated in the covariant calculations of the form factor with the symmetric and non-symmetric vertices in LFCQM.
The corresponding BS amplitude models a lightly bound heavy-light system with respect to the light quark mass. The infrared dynamics
appears to be important for the $D$ meson, as the sensitivity to changes of the constituent quark mass by about 10\% suggests.
This small modification produces a sensible change in the nearby unphysical continuum threshold, as the mass of the $D$ is about
the sum of the constituent quark masses. Therefore, the condition $m_1+m_2>M$ is barely satisfied (see, {\em e.g.}, the denominator
of Eq.~(\ref{ffactor+a})).

In summary, our study of the model dependence of elastic form
factors for pseudoscalar mesons in the light and heavy-light sectors
suggests a separation of the ultraviolet and infrared physics. The
ultraviolet properties of QCD dominate the form factor of light
pseudoscalars, whereas infrared physics and details of quark
confinement appear to be important for the space-like form factor of
heavy mesons below $q^2=10$~GeV$^2$. Hence, the electromagnetic form
factors of heavy-light systems, such as the $D$ and $B$ mesons,
provide a valuable tool in the effort to investigate the
nonperturbative physics of confinement, in contrast with light
pseudoscalar form factors insensitive to the details of infrared
physics even at low momentum transfers  once the weak decay
constants are fixed to the experimental values.

\vspace*{-5mm}

\acknowledgments
J.~P.~B.~C.~M. and T.~F. thank the LPNHE for the kind
hospitality during their visits when part of this work was developed. B.~E. is
grateful to Lauro Tomio for a pleasant stay at the
Instituto de F\'isica Te\'orica and to the  Universidade Cruzeiro do Sul for the welcoming
atmosphere; he also enjoyed valuable discussions with C.~D. Roberts.
This work was supported by the Department of Energy,
Office of Nuclear Physics, contract  no.~DE-AC02-06CH11357. We acknowledge  bilateral funding from the Centre National de la Recherche
Scientifique (CNRS) and Funda\c c\~ao de Amparo
\`a Pesquisa do Estado de S\~ao Paulo FAPESP)
under grant no.~06/50343-8 as well
as partial support from the Conselho Nacional de Desenvolvimento Cient\'\i fico e Tecnol\'ogico (CNPq).

\end{document}